\documentstyle[11pt,aaspp4]{article}

\def\teff{$T_{\scriptsize {\mbox{eff}}}$}
\def\kms{km s$^{-1}$}
\def\hb{H$\beta$}
\def\la{$\lambda$}
\def\mdot{$M_{\odot}$}
\def\he2{\ion{He}{2}}

\lefthead{Edmonds et al.}
\righthead{CVs and a He WD in NGC 6397}

\begin{document}

\title{Cataclysmic Variables and a Candidate Helium White Dwarf in the
Globular Cluster NGC 6397 \altaffilmark{1}}

\author{Peter D. Edmonds \& Jonathan E. Grindlay}
\affil{Harvard College Observatory, 60 Garden Street, Cambridge, MA 02138}
\authoremail{pedmonds@cfa.harvard.edu,josh@cfa.harvard.edu}

\altaffiltext{1}{Based on observations with the NASA/ESA Hubble Space
                 Telescope obtained at the Space Telescope Science 
                 Institute, which is operated by Association of
                 Universities for Research in Astronomy, Incorporated,
                 under NASA contract NAS 5-26555}

\author{Adrienne Cool}
\affil{Department of Physics and Astronomy, San Francisco State
University\\1600 Holloway Avenue, San Francisco, CA 94132}
\authoremail{cool@sfsu.edu}

\author{Haldan Cohn \& Phyllis Lugger}
\affil{Department of Astronomy, Indiana University, Swain West\\
Bloomington, IN, 47405}
\authoremail{cohn@indiana.edu,lugger@indiana.edu}

\and

\author{Charles Bailyn}
\affil{Department of Astronomy, Yale University, P.O. Box 208101
New Haven, CT, 06520-8101}
\authoremail{bailyn@astro.yale.edu}

\begin{abstract}

We have used the Hubble Space Telescope ({\it HST}) and the Faint Object
Spectrograph (FOS) to study faint UV stars in the core of the nearby
globular cluster NGC 6397.  We confirm the presence of a 4th cataclysmic
variable (CV) in NGC 6397 (hereafter CV 4), and we use the photometry of
Cool et al. (1998) to present evidence that CVs 1--4 all have faint disks
and probably low accretion rates.  By combining these results with new UV
spectra of CV 1 and the published spectra of Grindlay et al. (1995) we
present new evidence that CVs 1--3 may be DQ Her systems, as originally
suggested by Grindlay et al. (1995), and we show that CV 4 may either be a
dwarf nova or another magnetic system.  Another possibility is that the CVs
could be old novae in hibernation between nova eruptions (Shara et
al. 1986).

We also present the first spectrum of a member of a new class of UV bright
stars in NGC 6397 (Cool et al. 1998). These faint, hot stars do not vary,
unlike the CVs, and are thus denoted as ``non-flickerers'' (NFs).  Like the
CVs, their spatial concentration is strongly concentrated toward the
cluster center. Using detailed comparisons with stellar atmosphere models
we have determined log g = 6.25 $\pm$ 1.0, and \teff\ = 17,500 $\pm$ 5,000
K for this NF.  Using these line parameters and the luminosity of the NF we
show that the NF spectrum is consistent with a helium WD having a mass of
$\sim$0.25\mdot\ and an age between 0.1 and 0.5 Gyr (depending on the
models used). The NF spectrum appears to be significantly Doppler shifted
from the expected wavelength, suggesting the presence of a dark, massive
companion, probably a carbon-oxygen (CO) WD.

\end{abstract}


\section{Introduction} \label{sec.int}

Binaries have long been thought to have a crucial impact on globular
cluster dynamics and evolution (Hut et al. 1992), but only recently
(especially with the use of {\it HST}) have large numbers of them been
found in globular cluster cores where they are expected to act as the
central energy source that drives cluster expansion.  Discoveries of binary
millisecond pulsars (e.g. Manchester et al. 1991) and multiple
low-luminosity X-ray sources (Hertz \& Grindlay 1983) have recently been
supplemented by discoveries of large numbers of eclipsing binaries in
globulars (e.g. 47 Tuc; Edmonds et al. 1996 and Kaluzny et al. 1998) and a
significant population of main sequence--main sequence binaries in NGC 6752
(Rubenstein and Bailyn 1997).  Another recent breakthrough has been the
discovery of cataclysmic variables (CVs) in the cores of globular clusters,
using either dwarf nova (DN) outbursts in M5 (Oosterhoff 1941), 47 Tuc
(Paresce \& De Marchi 1994) and NGC 6624 (Shara, Zurek \& Rich 1996), UV
excess to recover an old nova in M80 (Shara and Drissen 1995) or
narrow-band H$\alpha$ emission. Using the latter technique 3 CVs have been
reported in NGC 6397 by Cool et al. (1995) and Grindlay et al. (1995;
hereafter GC95), and a fourth CV candidate by Cool et al. (1998; hereafter
CG98). Also, 2 probable CVs have been reported in NGC 6752 by Bailyn et
al. (1996). These CVs appear to be the long-sought optical counterparts of
the low-luminosity X-ray sources found in globular cluster cores. In
particular, the 3 brightest optical emission line objects in NGC 6397
(GC95) are the probable counterparts of the 3 brightest X-ray sources found
by Cool et al. (1993; see also Cool et al. 1995).

Observations of CVs in clusters can be used for a variety of studies
including: (1) CV formation and evolution in low-metallicity environments,
(2) stellar interactions in high-density environments, and (3) cluster
dynamical evolution.  Probable formation mechanisms for globular cluster
CVs include tidal capture and exchange collisions between main sequence
(MS) stars and white dwarfs (WDs), complementing studies of the MS star -
MS star interactions that produce blue stragglers. Since these formation
mechanisms differ from those for field CVs, and the stellar environment is
different, it would not be surprising to find systematic differences
between globular cluster and field CVs. In particular, GC95 and Grindlay
(1996) have suggested that the CVs in NGC 6397 might have a much higher
percentage of magnetic WDs than field CVs.  The dense, collapsed core of
NGC 6397 is a prime region to study the effects of stellar interactions
because its high central density makes interaction rates large and its
relative proximity at 2.2 kpc makes it possible to probe the core with high
spatial resolution (CG98).

This paper presents new {\it HST}/FOS spectra of CV 1 (from GC95) and CV 4
(from CG98) in NGC 6397.  By combining all available spectra for CVs 1--4
with the photometry of CG98 and comparing with field CVs we show that the
CV disks are consistent with those of faint quiescent DNe, given their
expected periods, but that their \he2\ \la4686 lines are unusually strong
for DNe (such systems would probably have long recurrence times between
outbursts). Instead, we argue that CVs 1--3 may be magnetic CVs (or perhaps
old novae), and that CV 4 is either a low accretion rate DN or a magnetic
CV.  It is possible that these objects may even be quiescent LMXBs,
although this is unlikely based on detailed comparisons with the x-ray and
optical properties (Grindlay 1996, 1998).  In any case we present good
evidence that there {\it are} systematic differences between populations of
globular cluster and field CVs.

Along with the 3 previously known classes of UV bright stars in NGC 6397
(blue stragglers, CVs and WDs), CG98 have discovered another class of UV
bright stars. Three faint, hot stars have been found within only $\sim$16$''$
of the cluster center, all of them non-variable (unlike the flickering
CVs).  CG98 have argued that these non-flickering (NF) stars are unlikely
to be CVs, ``normal'' CO WDs (recently evolved from single red giants),
extended horizontal branch stars or field stars, but instead that they are
good candidates for low-mass helium WDs.

Helium WDs have masses $\lesssim$ 0.49\mdot, and in the field are usually
found in binary systems containing either another WD or a neutron star
(Marsh, Dhillon \& Duck 1995, and Rappaport et al. 1995). These double
degenerates are thought to form by Roche lobe overflow (and usually common
envelope events) in primordial binaries containing red giants, if He
ignition in the red giant core, a proto helium WD, is avoided (Iben,
Tutukov and Yungelson 1997 discuss detailed formation scenarios). Several
low-mass WDs have been found or inferred in open and globular clusters
including the helium WD -- red giant binary S1040 in M67 (Landsman et
al. 1997), and the ultra--short period X-ray binary systems 4U 1820-30 in
NGC 6624 (Anderson et al. 1997) and Star S in NGC 6712 (Anderson et
al. 1993). S1040 in M67 probably formed after a subgiant underwent Roche
lobe overflow in a primordial binary (Landsman et al. 1997), but in denser
globular clusters, primordial binary evolution may be less important than
interactions involving subgiants or red giants. For example, red giant/WD
or red giant/MS star direct collisions should cause a helium WD to be left
behind in a binary system (Davies, Benz \& Hills 1991). Systems such as 4U
1820-30 and Star S probably result from either neutron star/red giant
collisions (Verbunt 1987) or neutron star/MS star capture and delayed mass
transfer (Bailyn \& Grindlay 1987).

Here, we report the first spectrum of one of the NGC 6397 NFs. The lack of
emission lines in the spectrum provides extra evidence against the CV
possibility and the log g value presented here argues against a CO WD or
extreme horizontal branch identification.  By comparing with published
model atmospheres we determine log g and \teff\ for the NF and we then
compare these parameters (along with the luminosity) with WD evolutionary
models to show that a low-mass helium WD is, indeed, a plausible
explanation for the NF. We also present evidence that the NF spectrum is
significantly Doppler shifted from the expected wavelength, suggesting that
the NF is in a binary system with a massive dark companion.

\section{Observations and Analysis} \label{sec.obs}

\subsection{CV 1} \label{sec.cv1}

Spectroscopic observations with {\it HST} were made of CV 1, the brightest
CV of the 3 studied by GC95, on October 2nd, 1996. Ultraviolet and optical
observations were obtained with the FOS/PRISM, giving a spectral coverage
from 1850--8950\AA\ with variable dispersion and resolution decreasing from
the blue to the red. Ultraviolet observations were obtained with the
FOS/G160L grating, with a spectral coverage from 1150--2510\AA\ (at a
spectral resolution of 9.2\AA), including a small contribution from
second-order geocoronal Ly$\alpha$ near the red end. The data were
originally reduced using the normal pipeline processing but were then
recalibrated using updated flat-fields and inverse sensitivity files
(removing some flat-field features at around the 5\% level).

\begin{figure}[htb]
\vbox to9cm{\rule{0pt}{9cm}}
\includegraphics{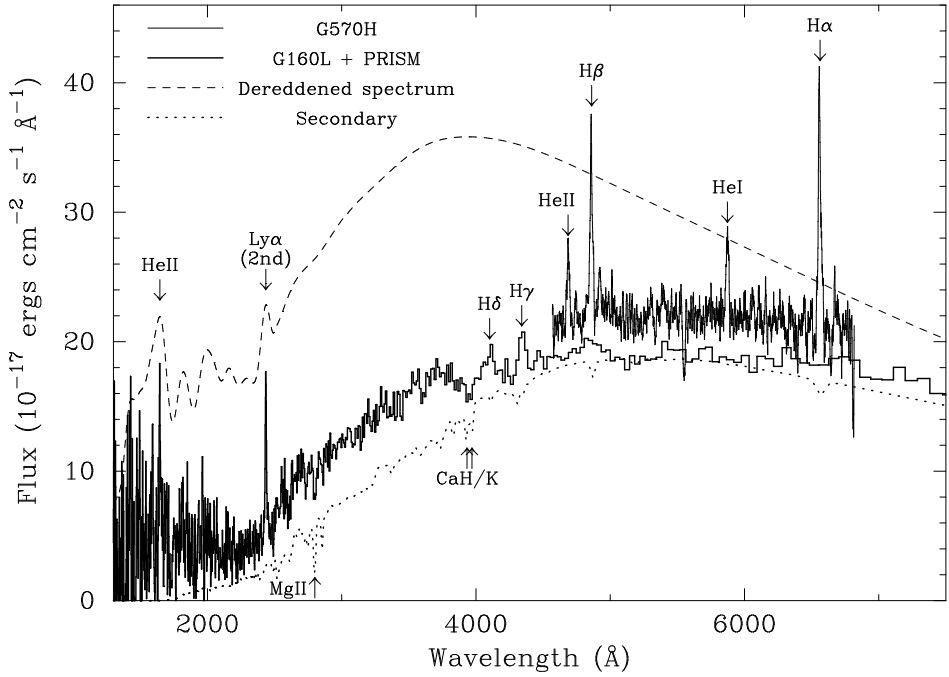}
\caption{A plot of all the FOS data available for CV 1: the full G160L
spectrum and the PRISM spectrum from 2510\AA\ red-ward plus the red G570H
spectrum from Cycle 4. Also shown is the dereddened smoothed continuum and
the estimated Kurucz spectrum of the secondary (see text).}
\label{fig.allcv1}
\end{figure}

Figure \ref{fig.allcv1} summarizes all of the FOS data available for CV 1.
The full G160L spectrum is shown and then the PRISM spectrum from 2510\AA\
red-ward, along with the red G570H spectrum (with a resolution of 4.5 \AA)
from Cycle 4.  No scaling was used for either the G160L or PRISM spectra,
showing good self-consistency in the spectral calibration and the
photometric states of (variable) CV 1 between the two separate
observations.  Despite the low spectral resolution of the PRISM in the
blue, the H$\gamma$ and H$\delta$ emission lines are visible, along with
some absorption lines from the secondary (for example MgII \la2800 and the
Ca H/K doublet). Since NGC 6397 lies close to the galactic plane its
reddening is significant and therefore Figure \ref{fig.allcv1} also shows a
dereddened version of the smoothed PRISM spectrum (using $E(B-V)=0.17$ from
Alcaino et al. 1997).

Using the WFPC2 photometry from CG98 we estimated the UV and optical
contribution from the secondary component in CV 1. We began by noting that
CV 1 is almost on the MS in the $V$ vs $V-I$ CMD, which implies a bright
secondary and relatively faint disk\footnote{Note that: (1) we use the term
``disk'' loosely for any hot, accretion related element of the binary
system, and (2) in general the total CV light minus the secondary will have
contributions from both the disk and WD, however the WD component is likely
to be negligible, particularly in the $B$ and $V$ passbands.} for this
system. Then, assuming that all of the flux in $I$ comes from the secondary
we used the position of the MS in CG98 to estimate $B$ and $V$ for the
secondary (showing the advantage of observing binaries in a cluster). We
then chose the reddened Kurucz atmosphere (log Z = --2.0) which best
matched the $BVI$ photometry for the secondary. Figure \ref{fig.allcv1}
shows the Kurucz stellar atmosphere with a dotted line.

\begin{figure}[b]
\vbox to1cm{\rule{0pt}{1cm}}
\includegraphics{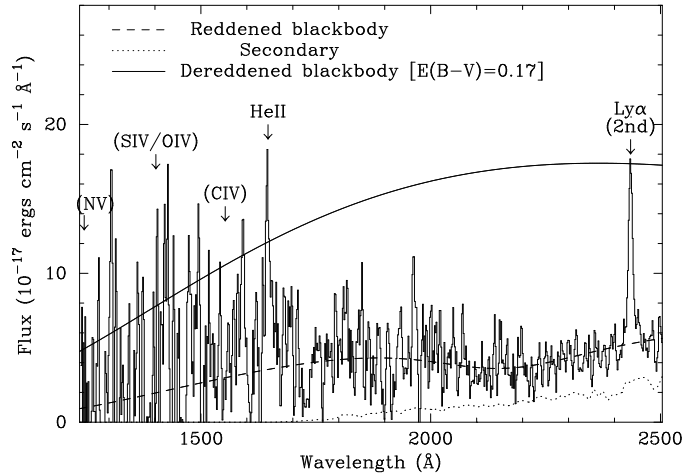}
\caption{A close--up of the G160L spectrum. Also shown are a reddened
blackbody fit to the UV spectrum (temperature = 12850 K), dereddened
spectra of this fit using three different values of $E(B-V)$ and the
estimated Kurucz spectrum of the secondary. \he2\ \la1640 and Ly $\alpha$
in 2nd order are the only 2 emission lines detected.} \label{fig.g160}
\end{figure}

A close--up of the G160L spectrum is shown in Figure \ref{fig.g160}. The
dashed line shows a reddened blackbody fit to the UV spectrum (temperature
= 12850 K, without removing the small contribution of the secondary) and
the solid line shows dereddening of this fit using $E(B-V)=0.17$. The
detection of Ly$\alpha$ in second order and a marginal (3.5$\sigma$)
detection of \he2\ \la1640 are labeled. No other UV lines are detected, and
we set 3$\sigma$ upper limits on the equivalent width (EW) of NV \la1246
(449\AA), SIV/OIV \la1402 (109\AA) and CIV \la1553 (58\AA).

The most striking feature of the G160L and PRISM spectra is the relatively
low UV flux. We have compared the measured flux distribution of CV 1 with
that of various subclasses of field CVs studied by Verbunt (1987) with
IUE. This comparison has limitations, as pointed out by the referee, since
Verbunt's sample may not include the full range of CV properties (including
highly magnetic WD primaries and a possible lack of disks in some DQ Hers).
However, the data set and analysis is homogeneous and covers most CV
classes, with the notable exception of AM Her systems (see below).

Verbunt (1987) determined the fluxes of CVs in 80\AA\ width bins centered
on 1460, 1800, 2140 and 2880 \AA. We first normalized each (quiescent)
system in Verbunt's study to have the same flux at 1460\AA\ and then
averaged the fluxes at other wavelengths over each class. We then
predicted, for each CV class, the average UV fluxes expected for $V$=18.27,
the magnitude of CV 1 ($V$ magnitudes are also given in Verbunt 1987). We
found reddened fluxes of $F_{1460} = 4.3 \times 10^{-16}$ergs
cm$^{-2}$s$^{-1}$\AA$^{-1}$ (DQ Hers), $4.0 \times 10^{-16}$ (U Gems) and
$4.5 \times 10^{-16}$ (Z Cams), and even brighter for other CV classes (for
example $F_{1460} = 1.2 \times 10^{-15}$ for old novae). These fluxes are
all over ten times brighter than the observed (reddened) $F_{1460}$ for CV
1 (see Figures \ref{fig.allcv1} and \ref{fig.g160}). However, scaling by
the $V$ magnitude of the system has limited usefulness because the
secondary is more massive and brighter than most field CVs.

\begin{figure}[htb]
\vbox to5cm{\rule{0pt}{5cm}}
\includegraphics{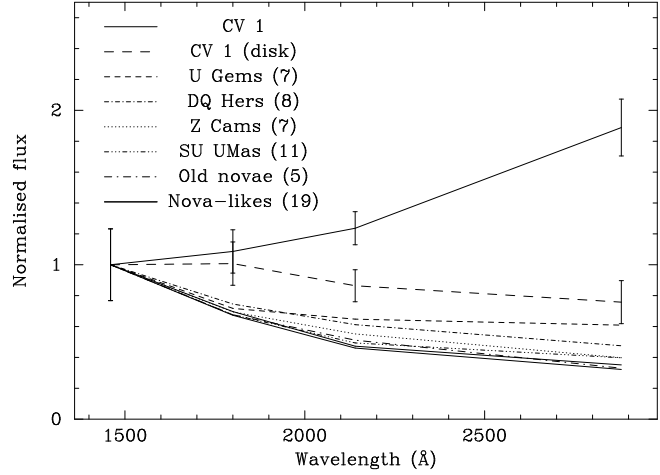}
\caption{The dereddened UV fluxes for CV 1 and the CV 1 disk (total
continuum minus the Kurucz spectrum of the secondary) along with fluxes for
various CV classes (number of systems given in parentheses).  The fluxes
have been normalized by the flux at 1460 \AA. The error bars combine
uncertainties in the continuum level estimation, the reddening and the
companion star estimate.} \label{fig.uv} 
\end{figure}

A better comparison is to calculate relative fluxes in the UV for CV 1,
where the contribution from the secondary is less important. The dereddened
UV fluxes for CV 1 are shown in Figure \ref{fig.uv} along with fluxes for
various CV classes (the error bars combine uncertainties in the continuum
level estimation and the reddening).  Not surprisingly, the reddest flux
distributions are found for quiescent dwarf novae (DNe), characterized by
low accretion rates, and DQ Her systems, with inner disks truncated by the
magnetic field of the WD.  Clearly, CV 1 has a much redder UV flux
distribution than the field CVs in Verbunt's sample. Figure \ref{fig.uv}
also plots the flux distribution for the CV 1 disk after subtracting off
the secondary and shows that the disk is marginally redder, given the large
errors, than all classes of field CV (without the secondaries
subtracted). 

\subsection{CV 4} \label{sec.cv4}

\begin{figure}[htb]
\vbox to5cm{\rule{0pt}{5cm}} 
\includegraphics{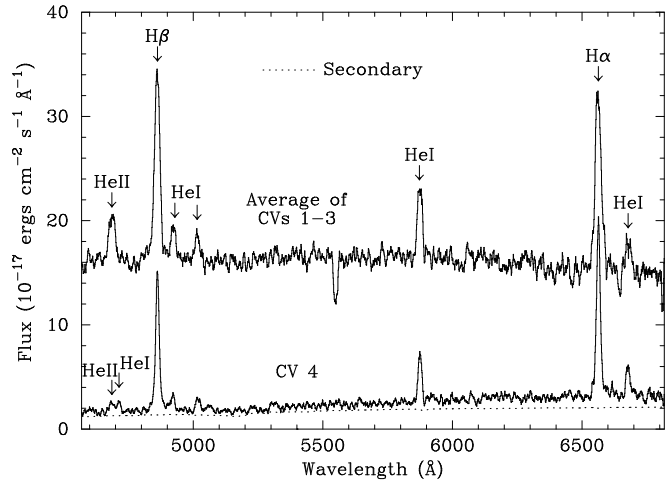} 
\caption{The diffuse light--corrected G570H spectrum of CV 4 and the
average spectra of CVs 1--3 from GC95, where the continua of CVs 2 and 3
were normalized to that of CV 1 (the feature at 5550 \AA\ is an
instrumental artifact). The dotted line shows the estimated contribution of
the secondary for CV 4. The emission lines detected are labeled.  Note the
relative weakness of the \he2\ \la4686 line for CV 4 compared to CVs 1--3,
and the detection of HeI \la4713.} \label{fig.cv4}
\end{figure}

We have obtained an optical spectrum of a fourth CV candidate discovered
near the center of NGC 6397 (CG98). Figure \ref{fig.cv4} shows the G570H
spectrum of this star, after correcting for diffuse light (from a
combination of extended PSF halos from bright stars such as giants and
diffuse light from faint cluster stars). For comparison, we also show an
average of the spectra of CVs 1--3 from GC95, where the continua of CVs 2
and 3 were normalized to that of CV 1.  Strong Balmer emission lines are
present in the spectrum of the new CV candidate along with several HeI
lines and \he2\ \la4686. The spectrum confirms that this star is a CV, as
suggested by its UV excess and flickering (CG98). The relative weakness of
the \he2\ line is the most obvious difference between CV 4 and the average
spectrum of CVs 1--3 (see a quantitative analysis below). The spectrum of
CV 4 has the highest signal to noise (S/N) ratio of the 4 CV spectra,
enabling easy detection of relatively weak lines such as HeI \la4713. The
dotted line shows the estimated contribution of the secondary, using the
procedure outlined above for CV 1.

To improve the spectra, we applied boxcar smoothing over 7 channels (this
improved the S/N ratio per independent channel by a factor of 2.6 at \hb,
but in simulations had a negligible effect on measured line parameters).
We found that a Voigt profile gave optimal fits to the emission lines,
after also experimenting with Gaussians and Lorentzians.

Before deriving EWs (see Section \ref{sec.he2}) and integrated magnitudes
from the spectrum of this faint star ($V$=20.81), we considered 2 possible
additional sources of light in the 0.43$''$ aperture used: (1) possible
light from individual neighboring stars and (2) the diffuse light
contribution.  By examining the high S/N WFPC2 images of CG98 we found that
light from neighbors is negligible, since CV 4 has no measurable companions
$< 0.93''$ away, but that diffuse light is an important factor for this
star.  To measure the spectral contribution of the diffuse light we used
Kurucz (1993) stellar atmospheres to fit a single temperature model to the
diffuse light components in $B$, $V$ and $I$ derived by CG98. A 6,000 K
Kurucz model gave an excellent fit and showed that relatively hot stars
dominated the diffuse light at the position of CV 4. Around 45\% of the
flux at 5500\AA\ comes from the diffuse light showing the importance of the
above correction for accurate EWs. By comparing the CV 4 spectrum with the
Kurucz spectrum of the secondary (see Figure \ref{fig.cv4}) we see some
evidence for a small residual red component.  This component is too red to
be caused by the CV disk (or WD), and is perhaps caused by earthshine (a
red component also appears, for data near the earth limb, to be present in
the spectrum of the NF; see below).

\subsection{\he2\ \la4686 line strength} \label{sec.he2}

An important diagnostic for CVs are the line fluxes, especially the
relative strength of \he2\ \la4686.  Table 1 gives values of the
wavelength, line flux, EW and FWHM (Gaussian and Lorentzian) of the \hb,
H$\alpha$, HeI \la5876 and \he2\ \la4686 emission lines for CVs 1--4
(1$\sigma$ errors are quoted). We have not undertaken a complete analysis
for CVs 1--3, which requires taking account of light contaminants in the
FOS aperture (CV 2, for example, has considerable contamination from a
bright neighboring star). However, the line ratio between \he2\ \la4686 and
\hb\ will be much less sensitive to contamination errors than the EWs and
so we derive these ratios from the uncorrected spectra, after taking into
account blends from HeI \la4713. Since multiline fitting in
IRAF\footnote{IRAF is distributed by the National Optical Astronomy
Observatories, which are operated by the Association of Universities for
Research in Astronomy, Inc., under cooperative agreement with the National
Science Foundation.}/SPLOT was unable to separate the heavily blended \he2\
\la4686 and HeI \la4713 lines in CVs 1--3, we used the relative strength of
HeI \la4713 compared to HeI \la5876 for CV 4 (where the blend from the weak
\he2\ \la4686 line was small; see Figure \ref{fig.cv4}) to remove the HeI
\la4713 line from these systems.

The \he2\ \la4686/\hb\ line ratio for CV 4 is only 0.07 $\pm$ 0.01, compared
to 0.32 $\pm$ 0.04, 0.34 $\pm$ 0.05 and 0.25 $\pm$ 0.03 for CVs 1--3
respectively (these more accurate determination replace those given in
GC95). Two other notable differences are that the lines of CV 4 are
narrower (smaller FWHMs) than those of CVs 1--3 and the EWs of most of the
CV 4 lines are greater than those of CVs 1--3. This latter result is part
of a very clear trend that the EWs increase going from CV 1 to CV 4, mainly
because the secondaries (which dominate the optical flux) get fainter,
without significant changes in the line fluxes. For example, comparing CV 4
with CV 1, the ratio of EWs of \hb\ is 8.8, with a factor of 0.08
difference in the continuum levels and only a factor of 0.7 difference in
line fluxes.

To compare the measured \he2\ \la4686/\hb\ line ratios with those of field
CVs we used the emission line data of Williams (1983) and Echevarria (1988)
for various CV classes (after confirming the CV classification using Ritter
\& Kolb 1998).  Because \he2\ \la4686 is often quite a weak line and it is
highly variable only a subset of the CVs in Echevarria's sample include
\he2\ \la4686, but even with this relatively small sample, some clear
trends emerge. The average \he2\ \la4686/\hb\ line ratios for the
non-magnetic CV classes are 0.16 $\pm$ 0.05 (17 DNe), 0.94 $\pm$ 0.50 (8
old novae) and 0.76$\pm$ 0.51 (13 nova--likes, excluding magnetic
systems). These values are probably overestimates (although we lack
information about upper limits), especially for DNe where only 17 out of 39
systems in the above sample have measurable \he2\ \la4686 (the completeness
for the other systems is 8/10 for old novae and 13/17 for nova--likes). For
magnetic systems we used the classifications by Patterson (1994) and Ritter
\& Kolb (1998) and the spectra of Williams (1983) to calculate an average
\he2\ \la4686/\hb\ line ratio of 0.57 $\pm$ 0.46 (measurable \he2\ \la4686
for 7/8 DQ Her systems) and 0.59 $\pm$ 0.21 (4/4 AM Her systems). The high
accretion rate novae and nova--likes can have relatively strong \he2\
\la4686 (probably because of very hot inner disks), while the magnetic
systems also usually have strong \he2\ \la4686, thought to be because the
hot accretion ``curtains'' along the magnetic field lines of the WD are
directly visible. Only the lower accretion rate DNe have relatively weak
\he2\ \la4686 compared to \hb.
 
\subsection{Disk brightness} \label{sec.disk}

For comparison with field CVs we have also estimated the brightness of the
disk in CVs 2--4, using the techniques described in Section \ref{sec.cv1}
to subtract off the secondary. Table 2 shows the total absolute magnitudes
of CVs 1--4, plus estimated absolute magnitudes and masses ($M_2$) of the
secondaries (from Bergbusch and Vandenberg 1992), and absolute magnitudes
of the disks ($M_V$ (disk)).  Since Smith and Dhillon (1998) have shown
that secondaries in CVs with orbital periods below $\sim$7 h, as likely
here (see discussion below) are typically not evolved, the mass estimates
for main sequence stars should be reasonable.  Note that, using this
technique, we found that the $V$ magnitude of the secondary in CV 1 was
0.02 mag {\it brighter} than the $V$ magnitude of the total system, clearly
an impossibility.  Variability is the likely explanation, plus errors will
also play a part; to investigate the sensitivity of these results to
overestimating the brightness of the secondary (and other errors) we made
the estimated $I$ magnitude of the secondary fainter by 0.1 mag and 0.2 mag
and rederived the expected $M_V$ (disk) (see Table 2).

We also show in Table 2, estimates of upper limits for periods of CVs 1--4,
using the relationship given in Warner (1995) between the mass of a
Roche-lobe filling secondary and the CV period, namely $M_2 = 0.065
P_{orb}^{5/4}$(h). Since this equation applies to Pop I stars and Pop II
stars have smaller radii for a given mass than Pop I stars (implying a
smaller period for Roche-lobe overflow), we have used the study by Stehle,
Kolb \& Ritter (1997) to estimate the period correction required for Pop II
stars ($<$ 1 h for all masses $<$ 0.9\mdot). As a guide to the validity of
this procedure for PopII stars we note that the two CV candidates in NGC
6752 (Bailyn et al. 1996), have absolute $V$ and $I$ magnitudes only a few
tenths of a magnitude different from CVs 2 and 3 in NGC 6397. The estimated
periods of 4.4 h and 3.8 h for CVs 2 and 3 compare well with the observed
periods of 5.1 h and 3.7 h for the NGC 6752 CV candidates.

\subsection{Non Flickerer}

The G570H spectrum of the NF is shown in Figure \ref{fig.allnf}. The blue
continuum and broad \hb\ line suggest that this star is a high gravity
object (see below). The spectrum suffers from both diffuse light
contamination and from having a much brighter ($\Delta$V = 3.5 mag)
neighbor only 0.3$''$ distant (a star near the MS turnoff). With perfect
pointing during the FOS observations we would simply be able to use the
high S/N WFPC2 images from CG98 to determine the exact amount of turnoff
star contamination. However, there are 0.08$''$ uncertainties in the
pointing of {\it HST} which can potentially make a significant difference
to the amount of light contamination.  There is also a source of variable
light, since the apparent flux of the star increases by about 15\% towards
the end of an orbit (probably due to increased scattered light or
earthshine as noted above).

Since it is difficult to estimate the ``clean'', uncontaminated spectrum of
this star we forced its continuum to equal that of a reddened Kurucz
atmosphere constrained by the $B$, $V$ and $I$ measurements of CG98.  A log
Z = --2.0 Kurucz spectrum was normalized so that the equivalent $V$
magnitude agreed with Cool's $V$ magnitude, and the temperature was varied
to give the best agreement at $B$ and $I$. The best fit spectrum had a
temperature of 22,000 $\pm$ 7,000 K, where the error is 1$\sigma$. Use of
this technique means that errors in the continuum determination are
dominated by errors in the photometry only (normalization from a separate
observation is valid because CG98 have shown that the star does not vary)
without incurring larger (unknown) errors by attempting a poorly
constrained independent measurement.

Having determined the continuum level, the shape of the \hb\ line for the
NF must be determined.  Since both the neighboring turnoff star and diffuse
light contributions have their own spectral components, estimates of these
aperture contaminants are required before the decontaminated \hb\ line
profile can be determined (the contribution of the neighboring turnoff star
is most important because its relatively hot temperature implies a
reasonably strong \hb\ line).  Therefore, we still made an approximate
determination of the telescope pointing during the FOS observations.  We
compared the flux in the FOS spectrum with the flux derived from a 0.43$''$
test aperture centered over the NF in the $V$ image from CG98 (thus
including all the aperture light, not just that from the NF). We first
subtracted away from the original spectrum the earthshine component derived
earlier ($V$=21.7), leaving only constant components. The equivalent $V$
magnitude of this ``constant'' FOS component is 19.44 mag, in excellent
agreement with the $V$ magnitude derived from the WFPC2 image (19.45).
Since the NF does not vary and the pointing of {\it HST} was effectively
constant during the FOS observations (no drifts above the $\sim$0.001$''$
level), the contamination from the turnoff star in the FOS aperture must be
approximately the same as in the test WFPC2 aperture. Using Cool's
photometry we decomposed the FOS spectrum into its separate components of
NF, turnoff star and diffuse light (all fitted by Kurucz atmospheres). The
resulting residual is shown in Figure \ref{fig.allnf} and the close
agreement with zero shows we made a reasonable, self--consistent
determination of the various aperture components. Since the residual is so
red, its contribution to the NF \hb\ line profile should be negligible (the
source of this residual may be incomplete removal of the earthshine
component mentioned earlier).

As noted above, the high temperature and the broad \hb\ line appear
consistent with the hypothesis that this star is a high gravity object such
as a WD. To test this hypothesis we used the pure hydrogen atmosphere
models of Wesemael et al. (1980). Modern refinements of these models do not
offer any significant advantages in the analysis of this low S/N ratio
spectrum. The advantages of using these models are that they include
detailed line profiles and cover a large range in log g (4 $<$ log g $<$ 9)
and \teff\ (20,000 K $<$ \teff\ $<$ 150,000 K).

Since the only obvious line present (\hb\ in absorption) is broader than
the emission lines for CVs 1--4, and because the S/N ratio in this line is
low, we increased the length scale of the smoothing to 11 channels and
applied the same smoothing factor to the models. We resampled the Wesemael
line profiles to the resolution of the FOS data, applied smoothing and then
experimented with different model profile fits (Lorentzians and Voigt
profiles gave almost identical results).  To fit the line profile the ICFIT
algorithm within IRAF/SPLOT was used to fit the continuum and a Lorentzian
was used to fit the line profile. The line depth and EW were found to be
0.55 $\pm$ 0.05 and 23.9 $\pm$ 2.4 \AA\ respectively, where the errors are
a combination of random errors in the parameter measurements (determined by
Monte Carlo experiments) and systematic errors in the continuum and line
profile measurements.

For simplicity we defined each model \hb\ line with two parameters, the
line depth and the EW. We generated a complete grid of these two parameters
for all of the available Wesemael models, interpolated the log g = 1
spacing to log g = 0.25 and interpolated the temperature scale where
necessary (for some log g values fewer models were available). Since the
Wesemael models are for \teff\ $>$ 20,000 K, we used the Kurucz models
(with 4 $<$ log g $<$ 5 and \teff\ $<$ 20,000 K) to extend our line
depth/EW grid below 20,000 K down to 10,000 K. (Encouragingly, excellent
agreement was found between the Wesemael and Kurucz models in the overlap
region of \teff\ = 20,000 K, 4 $<$ log g $<$ 5; differences of at most 2\%
were found in the two line parameters.) We extrapolated the Kurucz models
to higher log g values (from log g = 5 to log g = 9), constrained by the
functional form of the Wesemael models for the line depth and EW at 20,000
K, using the same smoothing and interpolation as before (see comments below
about the validity of this extrapolation).

Finally, the measured line depth and EW were used to search for a solution
in log g and \teff. The optimal solution was found to be \teff\ = 17,500
$\pm$ 5,000 K and log g = 6.25 $\pm$ 1.0 (1$\sigma$ errors). The large
errors in the gravity and temperature are caused by the limited information
present in one noisy line, in particular the inability of the spectrum to
trace possible narrow line cores. Figure \ref{fig.contnf} shows the total
$\chi^2$ for the above solution as a function of \teff\ and log g. The
optimal solution with $\chi^2 = 0.08$ is marked (``L''). The first, second
and fourth contour levels correspond roughly to 1$\sigma$, 2$\sigma$ and
3$\sigma$ respectively. Figure \ref{fig.contnf} also shows the regions
where the Wesemael and Kurucz models have been used, and where they were
extrapolated. Although extrapolation may incur extra uncertainties, Figure
\ref{fig.contnf} shows that a Wesemael model having \teff\ = 20,000 and log
g = 6.5 (with $\chi^2 = 0.8$) is close to an optimal solution, without
requiring any extrapolation. The Kurucz models are useful because they show
that a low gravity solution is probably not consistent with the data
(besides independently checking the low temperature, low gravity Wesemael
models).

Since the errors for log g and \teff\ from the \hb\ line measurement are
considerable, we briefly discuss extra constraints on these two
parameters. First, we note from Figure \ref{fig.contnf} that relative to
the optimal solution (``L'') only high temperature/high gravity or low
temperature/low gravity solutions are allowed. The high gravity solution
with log g = 7.25 seems ruled out by the photometry of Cool, Sosin \& King
(1997), since the NFs clearly have lower gravities than log g = 7, the low
gravity limit of the models used. The low gravity solution with log g =
5.25 and \teff\ = 12,500 K has a temperature that is probably inconsistent
with the determination from the photometry of CG98 (22,000 $\pm$ 7,000 K).

To determine the wavelength of the \hb\ line we adopted a two-step
procedure: (1) we used the Lorentzian fit to the \hb\ line as a first-order
solution and then (2) used a spectral model (template) with fixed
continuum, line depth and EW [determined in (1)] but with variable
wavelength to refine the initial estimate.  Using template shifts of up to
10\AA\ in 0.1\AA\ steps we selected the shift which minimized the
difference between the template and measured spectrum. This procedure is
formally similar to a cross correlation, but it gives sub-pixel resolution
without polynomial fitting and allows us to easily weight the first-order
solution to optimize sensitivity to Doppler shifts. Using this procedure
the \hb\ line was determined to have a wavelength of 4865.8\AA, noticeably
redder than the laboratory value of 4861.3\AA\ [step (1) alone gave
4864.5\AA]. The random error given by IRAF/SPLOT is only 0.4\AA, so we
investigated the possibility of systematic effects. According to the {\it
HST} Data Handbook, the overall 1$\sigma$ random uncertainty is 0.7\AA\ (or
43 \kms) for the accuracy with which the wavelength scale is known in an
individual FOS spectrum. However, the possibility of filter grating wheel
(FGW) displacements means that a worst--case disagreement in wavelength of
over 4\AA\ is possible unless we have independent confirmation of the
wavelength scale (when the wavelength scale is fixed the overall 1$\sigma$
random uncertainty falls to 0.5\AA\ or 31 \kms).

Our only independent constraints on the wavelength scale are the G570H
spectra of CV 4 (obtained over 3 orbits), since no movement of the FGW was
made between the observations of the NF and CV 4. The wavelength of the
\hb\ line for CV 4 was determined to be 4861.8\AA. The 0.5\AA\ shift from
the laboratory value is likely to consist of the velocity of the emission
region in the binary, the cluster radial velocity of 21 \kms\ (0.34\AA),
orbital motion of the Earth around the Sun, and the motion of the telescope
itself (the last 2 effects being negligible given the resolution), plus
systematic effects because of the FGW position. While CV emission lines are
hardly ideal radial velocity standards in general, we believe that CV 4
provides a useful wavelength reference for several reasons: (1) the
measured wavelength for \hb\ is only 0.15\AA\ red-ward of the laboratory
value, taking the cluster radial velocity into account, (2) the CV 4 \hb\
wavelength measurement appears very stable since the 3 sub-observations for
the 3 separate orbits (separated in time by almost 2.5 hours, well over
half the expected period of CV 4) give wavelengths of 4861.7\AA, 4861.8\AA,
and 4862.0\AA, (3) the line is very symmetrical (unlike CVs 1 and 2) so
that the wavelength measurement is unlikely to have been skewed, and (4)
the wavelength of H$\alpha$ for CV 4 is 0.4\AA\ red-ward of the laboratory
value, in excellent agreement with the 0.5\AA\ red--shift for \hb. Also,
two of the 3 strongest HeI lines give consistent Doppler shifts red-ward of
the laboratory value: HeI 4921, shift = 0.35 $\pm$ 0.45\AA\ and HeI 5876,
shift = 0.32 $\pm$ 0.21\AA. Only HeI 6678 gives a different shift, --0.72
$\pm$ 0.25\AA\ relative to the laboratory value, however this line is over
1800\AA\ red-ward of \hb\ and therefore has less value as a velocity
reference.

Using the wavelength of \hb\ for CV 4 as a reference, the Doppler shift
of the NF was measured to be 247 $\pm$ 50 \kms. To derive the error we
added together in quadrature the random error (25 \kms), the systematic
instrumental error quoted above (31 \kms) and an estimate of other
systematic errors, including the use of CV 4 as a velocity reference
(0.5\AA\ or 31 \kms).

A closeup of the \hb\ line profile is shown in Figure \ref{fig.closenf}.
Note, in particular, the vertical lines showing the difference in central
wavelengths between the average NF and CV 4 \hb\ lines. The significance of
the apparent shift of \hb\ for the NF is also clearly shown by comparing
the two Lorentzians (the astrophysical significance of this 247 \kms\ shift
will be discussed in Section \ref{sec.hewd}). The variable flux but
constant wavelength of the \hb\ line for CV 4 is clearly shown by the 3
emission profiles (the apparently constant velocity will be discussed in a
future publication).

\section{Discussion} 

\subsection{Cataclysmic variables} \label{sec.discv}

One of the crucial advantages of studying cluster CVs, besides their known
distance, is the opportunity to study CVs with different metallicities from
those of field CVs (see la Dous, 1991 for examples of UV spectra of field
CVs). Although the bluest portion of the G160L spectrum is noisy, the
detection of \he2\ \la1640 coupled with the non-detection of CIV \la1550 is
probably a direct reflection of the low metallicity for NGC 6397. The lack
of obvious emission at MgII \la2800 may be another measure of the low
cluster metallicity, but we defer detailed spectral modeling incorporating
the cluster's low metallicity to a future publication.

We now discuss the significance of the UV distribution of CV 1, the disk
brightnesses of CVs 1--4 and their \he2\ line ratios. Regarding the UV flux
distribution, one obvious explanation for the redness of the disk explained
in Section \ref{sec.cv1}, given our sample size of one (for G160L data), is
an inclination effect. A second possible explanation is the relatively
large contribution from the bright secondary. This explanation seems
plausible in that: (1) our predicted period for CV 1 is 5.1 h, while the
average period of the 8 DQ Hers in Verbunt's sample is only 3.8 h (also,
PopII secondaries tend to be brighter than PopI secondaries at a given mass
- Stehle, Kolb \& Ritter 1997), and (2) if we replaced CV 1 by any one of
the other CVs the flux distribution would be much bluer. For example, while
the estimated secondary for CV 2 is $\sim$1.3 mag fainter than the
secondary for CV 1, the disk in CV 2 is {\it brighter} (in $U_{336}$) than
the disk in CV 1.

If we then ignore the contribution of the secondary (see Figure
\ref{fig.uv}) a simple interpretation of the flux distribution for CV 1
compared to the field CVs is that it has either: (1) a relatively faint
and/or cool disk because of a low accretion rate, or (2) a WD with a
moderately strong magnetic field. As noted earlier, AM Her systems (with
strong magnetic fields) are not included in Verbunt's sample. These
systems, lacking disks, can be quite red, however the likely periods of the
NGC 6397 CVs are longer than most AM Her systems (see Ritter \& Kolb
1998). Also, the line emission of the NGC 6397 CVs differs from field AM
Her systems (see below).

To investigate (1) we compared $M_V$ (disk) and the CV periods from Table 2
with Figures 3.5 and 9.8 of Warner (1995), plots of $M_V$ (disk) versus
orbital period for {\it non}-magnetic field CVs. (The two principal error
sources for field CVs are estimates of the secondary component and distance
estimates, errors that are considerably reduced by studying cluster
CVs). Clearly, the NGC 6397 CVs fall near the faint limit for CV disks when
their expected period is considered. Among field CVs only faint DNe in
quiescence have disks this faint, and these systems have much longer
recurrence times for outbursts than brighter DNe (Warner 1987), possibly
explaining the observed paucity of DN outbursts in globular clusters (Shara
et al. 1996). Generally CVs with periods above the period gap have disks
that are brighter than $M_V \sim 8$. Indeed, a large number of
observational and theoretical studies have concluded that CVs with periods
above the period gap generally have high accretion rates, with
correspondingly bright disks while CVs with periods below the period gap
have lower accretion rates with fainter disks (Patterson 1984, di Stefano
\& Rappaport 1994 and Stehle, Kolb \& Ritter 1997, the latter two studies
specifically for PopII CVs).

Another way to emphasize the unusually faint disks of the NGC 6397 CVs is
to compare globular cluster and open cluster CVs (minimizing uncertainties
in $M_V$ (disk) estimates).  There are now 6 probable CVs in globular
clusters with well-measured $VI$ magnitudes {\it all} lying at most
$\sim$0.2 magnitudes away from the MS in the $V$ vs $V-I$ CMD (4 in NGC
6397 and 2 in NGC 6752). We contrast this result with the 3 CVs detected in
open clusters that have quiescent $V-I$ colors blue-ward of the MS by
$\gtrsim 0.6$ mag (M67; Gilliland et al. 1991), $\sim$0.7 mag and $\sim$1.0
mag (both NGC 6791; Kaluzny et al. 1997). Since M67 and NGC 6791 are much
less dense and dynamically evolved than NGC 6397, their CVs are expected to
have evolved from primordial binary systems, just as with field CVs. Hence,
the detection of a nova--like and Z Cam (relatively high accretion rate DN)
system in NGC 6791 is not surprising, since in any sample of field CVs
these are among the brightest, intrinsically, because of their high
accretion rate (resulting in blue $V-I$ colors). The much lower metallicity
of NGC 6397 compared to metal-rich field or open cluster CVs should, in
general, mean that the NGC 6397 CVs have {\it higher} accretion rates (for
given binary parameters), according to Stehle, Kolb \& Ritter (1997),
implying that even brighter disks should be present.

A possible explanation we have already suggested (GC95), is that these CVs
are mostly magnetic systems with truncated and thus relatively faint disks.
Our original suggestion was based on the relative strength of \he2\ \la4686
compared to \hb\ for CVs 1--3. To summarise the results in Section
\ref{sec.he2}, for the AM Her systems (with their strong magnetic fields),
a strong \he2\ \la4686 line is a well known feature (Warner 1995). However,
known DQ Hers show a large range in \he2\ \la4686/\hb\ line ratios with
values ranging from over one (V533 Her) to zero (AE Aqr). To discriminate
between magnetic and non-magnetic systems (and eliminate high accretion
systems like nova--likes), Silber (1992) has shown that line ratios of
\he2\ \la4686/\hb\ $>$ 0.4 and EW(H$\beta) >$ 20\AA\ are reasonable
signatures of magnetic systems.

By combining our knowledge of $M_V$ (disk) and the \he2\ line ratios of CVs
1--4, we can attempt identification of these systems.  With its weak \he2\
line and faint disk, CV 4 is a reasonable candidate for a quiescent DN
system.  Since there is remarkable similarity between the \he2\
\la4686/\hb\ line ratios for CVs 1--3 and also similarities in the Balmer
line fluxes themselves (see Table 1), CVs 1--3 may have very similar
properties, as originally suggested by GC95. They do not appear to be
recent old novae or nova--likes because of their faint disks (with extra
evidence from their \he2\ \la4686 line ratios), nor do they appear to be
DNe because they have moderately strong \he2\ lines.  The final option is
magnetic systems. CVs 1--3 do not have \he2\ ratios as large as AM Her
systems, but while their ratios are also weaker than an ``average'' DQ Her
system, there is considerable scatter for the DQ Her systems, as pointed
out above. For example, CVs 1--3 have \he2\ ratios stronger than 4 of the 8
DQ Her systems in the sample of Williams (1983), and CVs 1 and 2 only just
fail the magnetic criteria of Silber (1992).  To conclude, CVs 1--3 do not
appear to be DNe, but they could be DQ Her type systems.

Is it true that DQ Her systems tend to have disks as faint as those found
in CVs 1--4?  Extra information about DQ Her disks compared to other CV
classes comes from the continuum slopes in Williams (1983). An estimate of
the disk contribution (and temperature) relative to the secondary comes
from analyzing the continuum ratio between two different wavelengths
(equivalent to a color). The two most convenient continuum points for both
CVs 1--4 and the spectra of Williams (1983) are at \hb\ and
H$\alpha$. Figure \ref{fig.he2disk} shows plots of this continuum ratio
versus the \he2/\hb\ line ratio for several different classes of field CV
(from Williams 1983, updated by Ritter \& Kolb 1998) along with these
ratios for CVs 1--4. We examined the linear correlation between the
continuum and line ratios for the individual CV classes shown, finding a
linear correlation with absolute value $>$ 0.5 in 3 cases: DQ Her systems
(linear correlation = 0.77), nova--likes (--0.69) and old novae (0.57). The
correlation for DQ Hers implies that small \he2\ line ratios imply
continuum ratios of $\sim$1.0 (meaning that the secondary dominates unless
the disk is very cool). A notable element of Figure \ref{fig.he2disk} is
that all of CVs 1--4 lie close to the best fit line for DQ Hers, and all of
them have continuum ratios of $\sim$1.0. The dominance of the secondary for
CVs 1--4 is therefore exactly what we expect for field DQ Her systems with
weak \he2\ lines.

Using estimates of $V$ (disk) and distances for DQ Her systems (from
Patterson 1994) we have estimated $M_V$ (disk) for DQ Hers, neglecting
reddening (the $\sim 50\%$ errors in the distances dominate the errors) and
assuming that the systems are in their faint, low accretion state.  We
found a strong linear correlation (--0.83) between the \he2/\hb\ line ratio
and $M_V$ (disk), as shown in the upper panel of Figure \ref{fig.he2disk},
in the sense that higher \he2/\hb\ implies a brighter disk (consistent with
the bluer colors given above). We then used this correlation to {\it
predict} $M_V$ (disk) using the measured \he2/\hb\ line ratios for CVs 1--4
given in Section \ref{sec.cv4}. The results are shown in the final column
of Table 2, and agree nicely with the $M_V$ (disk) values given in Table 2
for CVs 1--4, especially given the large uncertainties in $M_V$ for the
field CVs.  (We excluded GK Per from the sample, with its $\sim$2 day
period, evolved companion and thus high accretion rate.  Including GK Per
gave a linear correlation of --0.59 between $M_V$ (disk) and \he2/\hb\ and
$M_V$ (disk) values $\sim$0.6 mag brighter.)  These results are the best
evidence that these cluster CVs are indeed largely DQ Her type systems.

A possible alternative to the DQ Her hypothesis is that some of the 6397
CVs are old novae (possibly in deep hibernation between outbursts; see
Shara et al. 1986). This explanation is consistent with the correlation
noted above between the continuum ratios and \he2/\hb\ line ratios for old
novae, plus the proximity of the 6397 CVs to the best-fit line for old
novae.  Although the hibernation scenario is difficult to test (and yet to
be confirmed), upcoming {\it HST} spectra of the probable old nova in M80
will provide a critical comparison with the spectra of the NGC 6397 CVs. It
is also possible that some of the cluster CVs fall in both categories,
since, for example, 3 of the DQ Hers in the sample from Williams (1983) are
also old novae.

\subsection{Helium White Dwarfs} \label{sec.hewd}

The photometry of the NFs (CG98) and the spectroscopy presented above place
important constraints on the possible properties and evolution of these
systems, which appear to be helium WDs.  From the spectroscopy we have
already determined \teff\ and log g and from the photometry we calculated
the luminosity of the NF (log($L/L_{\odot})=-0.6$) using the bolometric
corrections presented in Bergeron, Wesemael \& Beauchamp (1995). Armed with
these three quantities we have examined evolutionary models of WDs to
estimate masses and lifetimes for the NFs. We began by using the WD
evolutionary models presented by Bergeron et al. (1995). Models with log g
= 7 (the limit of Bergeron's models), \teff\ = 17,500 K and with thick
hydrogen layers have $M \approx 0.3 $\mdot. Extrapolating to log g = 6.25,
models should have $M \approx 0.2-0.25$\mdot. Using the study by Webbink
(1975) of the evolution of helium WDs in close binaries, we used the
measured log($L/L_{\odot}$) and \teff\ to derive a mass between 0.2 and
0.25 \mdot\ for the NF, with cooling ages for these models between $\sim 2
\times 10^8$ yr (0.25\mdot) and $\sim 5 \times 10^8$ yr (0.20\mdot).  Using
log g, \teff\ and log($L/L_{\odot}$), a similar mass is determined from the
He WD study of Althaus \& Benvenuto (1997), although they do not include a
hydrogen envelope.

Since the only significant line present in the NF is \hb, a hydrogen
envelope must be present, and therefore we have made comparisons with the
models of Benvenuto \& Althaus (1998) on the effects of hydrogen envelopes
on the structure and evolution of low-- and intermediate--mass WDs. They
computed the evolution of WDs with masses from 0.15 -- 0.5 \mdot, while
treating the mass of the hydrogen envelope as a free parameter within the
range $10^{-8} \leq M_H/M \leq 4 \times 10^{-3}$. Because only a
representative sample of the models are presented in Benvenuto \& Althaus
(1998), L. Althaus has kindly performed a dedicated search for models
consistent with our log g and \teff. Two good solutions were found for a
0.24\mdot\ model with $M_H/M = 1 \times 10^{-3}$ and a 0.235\mdot\ model
with $M_H/M = 5 \times 10^{-4}$, with both of these solutions having
log($L/L_{\odot})=-0.52$, in excellent agreement with the measured
luminosity.  These solutions are close to the valid bright limit of the
models of Benvenuto \& Althaus (1998), so the cooling age for these
solutions of $\sim 7 \times 10^7$ yr is highly uncertain (and likely to be
an underestimate) since Benvenuto \& Althaus (1998) did not attempt a
detailed treatment of the formation of helium WDs in binary systems.

A detailed treatment of the evolution of a 0.3 \mdot\ WD in a binary system
was given by Iben \& Tutukov (1986; hereafter IT86). Their model
experiences two hydrogen shell flashes before cooling to \teff\ $\sim$
16,000 K at an age of $\sim 1 \times 10^8$ yr (a comparable age to that
found by Benvenuto \& Althaus (1998), but smaller than Webbink's
values). Using their cooling curve, IT86 construct a number--luminosity
distribution function for helium WDs. This function has a similar slope to
that for CO WDs, but at most magnitudes is about a factor of 4 smaller. One
exception to this behavior is the region with $-1 \lesssim
$log$(L/L_{\odot}) \lesssim -1.7$. At log$(L/L_{\odot}) \sim -1$ the
predicted helium WD number-luminosity function shows a ``bump'' where it
increases to be roughly equal to the CO WD function. It then drops quickly
with decreasing luminosity by about an order of magnitude before returning
to the average value at log$(L/L_{\odot}) \sim -1.7$. This behavior is
caused by the very slow rate of decline in luminosity following the first
hydrogen shell flash, when t = $10^7 - 10^8$ yr. Although IT86 warn that a
complete theoretical distribution function requires superposition of
contributions from WDs of many different masses (each experiencing a
hydrogen shell flash at a different luminosity), this oscillatory behavior
may cause an enhancement in the number of helium WDs at relatively high
luminosity and a dearth for $\sim$1.5 mag below this. This could be
consistent with the observed CMD distribution in CG98, perhaps explaining
the lack of obvious low-mass WDs further down a cooling sequence,
particularly since log$(L/L_{\odot}) \sim -1$ is not far from the measured
luminosities for the two faintest helium WD candidates (eventually close
double degenerate systems should merge after a few Gyr, forming a more
massive WD). However, clearly more data from other clusters are needed to
test this hypothesis.

An obvious question remains: are the possible formation mechanisms listed
in Section \ref{sec.int} consistent with the likely significant red--shift
of the \hb\ line?  We reject the possibility of the red--shift being an
ejection velocity, since at 247 \kms, the star would have moved 2.2 pc in
only $10^4$ yr, and thus would be well outside the core. Gravitational
red--shift is likely to contribute only a small red--shift for this low-mass
object, since $K_R = 0.635 \times M/M_\odot \times R_\odot/R~$ \kms\ (Reid
1996), and using the mass and radius estimates from Benvenuto and Althaus'
models, $K_R \approx$ 3 \kms. Instead, we argue that the red--shift may be a
Doppler shift from the helium WD orbiting a massive but faint companion,
probably a WD. This is consistent with the primary formation mechanism
listed above. Also, a binary nature for this star would hardly be
surprising for dynamical reasons alone, since the detection of 3 NFs near
the center of NGC 6397 is good evidence of a binary origin for these stars
(through mass segregation), and most known low-mass WDs are in binary
systems. Finally, we note the possibility that these low-mass WDs may have
neutron star companions, likely to be binary millisecond pulsars (although
none have been found yet in NGC 6397) as found in helium WD/NS binary
systems in the field.

If we have measured an orbital Doppler shift its size is an important
consideration. Of the 8 double--degenerate systems listed in Iben et
al. (1997), all with a helium WD as the primary (the brighter component),
only 2 have $K > 150$ \kms. Little is known about the masses of the
secondaries in these systems, but models by Iben et al. (1997) show that $K
\approx 250$ \kms\ should be quite common for high-inclination systems with
a helium WD primary and a CO WD secondary (from Cool, Piotto \& King 1996
and CG98, NGC 6397 clearly has a large reservoir of $\sim 0.55$ \mdot\ CO
WDs, and it should also have many higher mass WDs). For more massive
secondaries such as neutron stars, larger values of $K$ can be found, for
example $K = 280$ \kms\ for the helium WD -- MSP binary J1012+5307 (van
Kerkwijk, Bergeron \& Kulkarni 1996).

Adopting a range of possible mass functions for a binary system containing
the NF and a unknown companion, and assuming $K$ = 247 \kms\ and helium WD
mass = 0.25 \mdot, we have derived expected periods for the system as a
function of orbital inclination ($i$).  Although the NF observations were
spread over 3 (96 minute) orbits, $\sim$42 min of spectra were taken on the
middle orbit, but only $\sim$10 min at the end of the first orbit and
$\sim$14 min at the start of the third. Therefore, we cannot rule out a
$\sim$4-5 h period, and for $i < 60 \arcdeg$ we require a WD companion with
mass $\gtrsim$ 0.7 \mdot, slightly higher than the mass of WDs currently
being produced in the cluster (low-mass MS stars are ruled out, consistent
with the photometry of CG98). Since we have measured only a lower limit for
$K$, we have also experimented with values that are 25\% higher than
observed, maintaining the other assumptions.  In this case a high mass WD
(mass $\gtrsim$ 1.1 \mdot) or neutron star companion is required (clearly
longer observations may provide powerful constraints on the mass of the
companion).

To determine whether 3 systems with ages of $\sim 1 - 5 \times 10^{8}$ yr
(IT86 and Webbink 1975) are likely to be present in NGC 6397, we used
expected encounter rates between various combinations of MS stars, WDs,
neutron stars and red giants (Davies \& Benz 1995) to make rough estimates
of expected numbers of binaries or merger products. The calculations by
Davies \& Benz (1995) are specifically for $\omega$ Cen and 47 Tuc - we
used their numbers for the denser cluster 47 Tuc and divided the
interaction rates by 5 to account for the smaller mass of NGC 6397 (using
cluster masses from Pryor \& Meylan, 1993). Assuming a lifetime of 4 Gyr
(Sandquist, Bolte, \& Hernquist 1997) for blue stragglers less than
$\sim$1.5 mag brighter than the MS turnoff (where most of the observed blue
stragglers are found) and assuming all MS star collisions result in
mergers, we used the calculations of Davies \& Benz (1995) to estimate that
75 blue stragglers should currently be found in NGC 6397 (from both 2- and
3-body interactions), comparing favorably with the 54 blue straggler
candidates found by Kaluzny (1997). An overestimate of total blue
stragglers is not surprising because merged stars with total masses at or
below the turnoff are, of course, not included in Kaluzny's sample. For
CVs, we assumed that half the collisions between WDs and MS stars result in
the formation of binaries (Davies \& Benz 1995), and used a conservative
limit on the mass ratio of 1 for stable mass transfer (Davies \& Benz
1995).  Then, assuming the average CV lifetime is 1.5 Gyr for CVs above the
period gap (Kolb \& Stehle 1996), we expect 6 CVs from 2- and 3-body
interactions, close to the observed number.  Finally, assuming (1) ages of
0.1 -- 0.5 Gyr for the observed helium WDs, (2) that all of the red giant
collisions with either neutron stars or WDs result in the formation of He
WDs (see Davies, Benz \& Hills 1991), and (3) that 3-body interactions
result in as many helium WDs as 2-body interactions, we predict 0.7 -- 3.5
bright helium WDs to currently be found in NGC 6397.

The reasonable agreement between observed and estimated numbers of blue
stragglers and CVs is comforting given the large uncertainties in the
interaction rates used, the influence of possible {\it differences} in
interaction rates between NGC 6397 and 47 Tuc (e.g. NGC 6397 is
core-collapsed and 47 Tuc may or may not be), and the effect of the unknown
binary content (Davies \& Benz 1995 assume 10\% binaries). Since the
lifetimes for blue stragglers are known reasonably well and the number of
observed systems is probably fairly complete, the possible differences
listed above must either be small or cancel each other out. We also have
reasonable agreement between the numbers of expected and observed helium WDs
when using Webbink's cooling ages. Although our assumptions may appear
somewhat generous, there may be other mechanisms for the formation of He
WDs: for example, red giant -- MS star collisions, binary -- binary
collisions and primordial binary evolution.

It is the expected lifetimes for the helium WDs that are one of the largest
sources of uncertainty in the above number estimates. The (unknown)
thickness of the hydrogen envelope can make a significant difference to the
lifetimes especially if hydrogen burning occurs. For example, a model with
$M = 0.3$\mdot\ and $M_H/M = 2 \times 10^{-3}$ (the thickest envelope
considered by Benvenuto \& Althaus (1998), with log g and \teff\ consistent
with our low-mass NF within the errors) has an age $\sim$50\% greater than
the 0.3\mdot\ model with $M_H/M = 1 \times 10^{-3}$. It has also been
suggested (Alberts et al. 1996) that the cooling time--scales of very
low-mass WDs (mass $<$ 0.25\mdot) can be considerably underestimated by the
traditional WD cooling curves of IT86 and others for higher mass WDs (mass
$>$ 0.3\mdot). Alberts et al. (1996) predict that hydrogen shell flashes do
not occur on WDs with mass $<$ 0.2\mdot, but that these WDs experience
long--lasting phases of hydrogen burning which significantly prolong their
lifetimes (however this difference in behavior may be caused by differences
in time resolution between the models of Alberts et al. and IT86). Finally,
Sarna, Antipova \& Muslimov (1998) find much greater cooling ages for their
0.16\mdot\ WD compared to IT86 because of differences in the formation
mechanisms. While in IT86 the donor star fills its Roche lobe when it is on
the red giant branch, forming a $\sim$0.3\mdot\ helium WD with a relatively
thin hydrogen envelope, the donor star in Sarna et al.'s calculations fills
its Roche lobe while it is evolving through the Hertzsprung gap, resulting
in a 0.16\mdot\ WD with a much thicker hydrogen envelope.

Since bright red giants have much shorter lifetimes than subgiants (and
perhaps limited cross-sections, despite larger radii, because of low
densities) it is possible that collisions involving subgiants or faint red
giants are much more efficient at producing helium WDs than collisions
involving brighter red giants.  This selection effect would preferentially
cause very low-mass ($\lesssim$0.25\mdot) WDs to form (since the red giant
core mass increases with increasing brightness), giving objects with longer
cooling times according to Webbink (1975) and Sarna et al. (1998), and
enhancing the number of helium WDs seen.

\section{Conclusion} \label{sec.con}

We summarize here the results for CVs 1--4: (1) a 4th CV candidate in NGC
6397 has been spectroscopically confirmed, (2) UV data for CV 1 implies
that it has a red disk when compared with field CVs, (3) the photometry of
CG98 combined with Kurucz spectra for the estimated secondaries provide
strong evidence that CVs 1--4 all have faint disks and probably low
accretion rates (consistent with faint quiescent DNe), (4) the \he2\
\la4686/\hb\ line ratios (together with the faint disks) imply that CVs
1--3 may be DQ Her systems, (5) the correlations between the \he2/\hb\ line
ratios for CVs 1--4 and both (a) the continuum ratios between \hb\ and
H$\alpha$ and (b) $M_V$ (disk) provide extra evidence that the 6397 CVs are
mainly DQ Her systems. This is consistent with the finding of Verbunt et
al. (1997) that CVs 1--3 could be DQ Her systems based on their X-ray and
optical fluxes. An alternative explanation is that some of the CVs are old
novae in hibernating phases between nova eruptions.

We conclude that there may be fundamental differences between populations
of globular cluster CVs and field/open cluster CVs, perhaps caused by the
different formation mechanism of tidal capture and exchange collisions or
perhaps because of the different environment in globular clusters.  One
possible explanation is that interactions cause stars to rotate more
quickly, resulting in stronger magnetic fields than in most field stars, as
suggested by GC95. Alternatively, Vandenberg, Larson \& De Propris (1998)
have suggested that rapidly rotating cores of giant stars in the metal poor
globular cluster M30 might reconcile this cluster's luminosity function
with stellar evolutionary theory. Similar problems exist in understanding
the luminosity function of NGC 6397, although further study of the theory
of Vandenberg et al. (1998) is needed. Another possibility is that magnetic
WDs are formed in globulars from differentially rotating cores in blue
stragglers (Grindlay 1996).

Prospects for further work on the CV data include modeling of the disk for
CV 1, power spectrum analysis of both the time series obtained by CG98 and
the sub-exposures for the FOS spectra, studies of line profile changes with
time and detailed spectral modeling incorporating the cluster's low
metallicity. A clear test of the hypothesis that most of the 4 CVs are DQ
Her systems is to search for a stable optical (or X-ray) period with $P <
P_{orb}$ (DQ Hers usually have $P \ll P_{orb}$; Patterson 1994). Because
the short FOS observations of the CVs are inadequate for this purpose, we
shall propose to obtain simultaneous spectra of CVs 1 and 2 using moderate
time-resolution ($\Delta\tau \sim$ 60s) spectroscopy in the blue (using
STIS with a long slit), to directly test the magnetic CV hypothesis and
place constraints on the hibernating nova scenario.

The results for the NF are: (1) using detailed comparisons with stellar
atmospheres from Wesemael et al. (1980) and Kurucz (1993) we have
determined log g = 6.25 $\pm$ 1.0, and \teff\ = 17,500 $\pm$ 5,000 K
(consistent with \teff\ = 22,000 $\pm$ 7,000 K using the photometry of
CG98), (2) by using these line parameters and the luminosity of the NF we
have shown that the NF spectrum is consistent with a helium WD having a
mass of $\sim$0.25\mdot\ and an age between 0.1 and 0.5 Gyr (depending on
the models used), and (3) the NF spectrum appears to be significantly
Doppler shifted from the expected wavelength, suggesting the presence of a
dark, massive companion. The low mass of the NF (and probably similar or
lower masses for the others) suggest that interactions between degenerate
stars and subgiants or faint red giants are more efficient at producing
helium WDs than interactions involving degenerate stars and brighter red
giants.

Although we have not yet made a rigorous attempt to find evidence for
velocity variability of the \hb\ line for the NF, the prospects from
subdividing this low S/N spectrum are poor, especially since almost two
thirds of the data for the NF was obtained over just one $\sim$42 min time
segment. Clearly, observations over a longer time are needed to confirm
that Doppler shift evidence presented above and to study radial velocity
variations. Use of STIS with the long slit would enable two NFs to be
observed simultaneously, along with many cluster stars providing an ideal
radial velocity reference. Observations in the blue would also give
excellent coverage of Balmer absorption lines (with the exception of
H$\alpha$), giving considerably more accurate line parameters, and helping
determine whether the luminosity difference between the bright NF and the
two fainter ones is mainly because of mass or age differences.
 
\acknowledgments

We thank R. Kurucz, L. Althaus and O. Benvenuto for models, B. Hansen,
R. Di Stefano and F. Wesemael for discussions and an anonymous referee for
helpful comments. This work was partially supported by NASA grants
NAG5-3808 and HST grant GO-06742 (PDE and JEG), and NASA grant NAG5-6404
(CDB).



\begin{deluxetable}{lrrrrr}
\tablecolumns{6}
\tablewidth{0pc}
\tablecaption{Emission line data}
\tablehead{
\colhead{} & \colhead{Wavelength} & \colhead{Flux \tablenotemark{a}} & 
\colhead{EW} & \colhead{Gaussian\tablenotemark{b}} & 
\colhead{Lorentzian\tablenotemark{b}}  \\
\colhead{} & \colhead{} & \colhead{} & 
\colhead{} & \colhead{FWHM} & \colhead{FWHM} \\
\colhead{} & \colhead{(\AA)} & \colhead{} & \colhead{(\AA)} & 
\colhead{(\AA)} & \colhead{(\AA)} \\
} 
\startdata

\cutinhead{CV 1}

\ion{He}{2}& 4687.04 $\pm$ 0.59 & 1.26 $\pm$ 0.11 &   5.9 $\pm$ 0.5 &
                21.0 $\pm$ 2.0 & 0.0 $\pm$ 2.7   \nl
H$\beta$  & 4858.09 $\pm$ 0.21 & 3.92 $\pm$ 0.16 &  18.0 $\pm$ 0.8 &
                3.8 $\pm$ 5.7 & 15.6 $\pm$ 1.6   \nl
He I      & 5872.91 $\pm$ 0.45 & 1.51 $\pm$ 0.14 &   6.9 $\pm$ 0.7 &
                21.6 $\pm$ 2.2 & 0.0 $\pm$ 2.7   \nl
H$\alpha$ & 6559.19 $\pm$ 0.30 & 4.40 $\pm$ 0.18 &  21.0 $\pm$ 0.9 &
                24.0 $\pm$ 1.2 & 0.0 $\pm$ 1.6   \nl
\cutinhead{CV 2}

He II     & 4688.37 $\pm$ 0.86 & 1.44 $\pm$ 0.16 &  11.2 $\pm$ 1.3 &
              13.9 $\pm$ 12.0 & 23.6 $\pm$ 7.1   \nl
H$\beta$  & 4862.47 $\pm$ 0.17 & 4.19 $\pm$ 0.15 &  31.9 $\pm$ 1.2 &
              20.8 $\pm$ 2.0 & 11.1 $\pm$ 2.1   \nl
He I      & 5873.98 $\pm$ 0.53 & 1.04 $\pm$ 0.10 &   7.6 $\pm$ 0.8 &
              21.7 $\pm$ 2.8 & 0.0 $\pm$ 3.2   \nl
H$\alpha$ & 6563.60 $\pm$ 0.39 & 3.94 $\pm$ 0.16 &  30.2 $\pm$ 1.3 &
              30.7 $\pm$ 1.5 & 2.1 $\pm$ 2.1   \nl
\cutinhead{CV 3}

He II     & 4685.71 $\pm$ 0.81 & 0.53 $\pm$ 0.05 &  15.2 $\pm$ 1.5 &
              26.2 $\pm$ 5.5 &10.6 $\pm$ 7.3   \nl
H$\beta$  & 4860.73 $\pm$ 0.12 & 2.07 $\pm$ 0.05 &  59.0 $\pm$ 1.6 &
              13.8 $\pm$ 0.9 &10.5 $\pm$ 0.8   \nl
He I      & 5874.59 $\pm$ 0.37 & 0.53 $\pm$ 0.03 &  13.6 $\pm$ 1.0 &
              18.6 $\pm$ 1.5 & 0.0 $\pm$ 1.6   \nl
H$\alpha$ & 6562.94 $\pm$ 0.24 & 2.81 $\pm$ 0.09 &  71.7 $\pm$ 2.8 &
               3.1 $\pm$ 4.1 &20.2 $\pm$ 1.4   \nl
\cutinhead{CV 4}

He II     & 4685.60 $\pm$ 0.67 & 0.20 $\pm$ 0.03 &  12.3 $\pm$ 1.8 &
              21.0 $\pm$ 3.6 & 0.0 $\pm$ 4.1   \nl
H$\beta$  & 4861.84 $\pm$ 0.05 & 2.73 $\pm$ 0.03 & 158.0 $\pm$ 2.7 &
               9.4 $\pm$ 0.5 & 8.1 $\pm$ 0.4   \nl
He I      & 5874.78 $\pm$ 0.18 & 0.79 $\pm$ 0.03 &  29.3 $\pm$ 1.4 &
              15.0 $\pm$ 0.8 & 0.0 $\pm$ 1.0   \nl
H$\alpha$ & 6563.27 $\pm$ 0.06 & 3.39 $\pm$ 0.04 & 106.7 $\pm$ 1.7 &
              11.9 $\pm$ 0.3 & 6.2 $\pm$ 0.4   \nl

\tablenotetext{a}{The flux is in units of $10^{-15}$ergs
cm$^{-2}$s$^{-1}$\AA$^{-1}$}

\tablenotetext{b}{The Gaussian and Lorentzian components combine to give
the Voigt profile}

\tablecomments{The 1$\sigma$ errors quoted here have been calculated by
Monte Carlo simulations within SPLOT based on the photon noise errors
only. Systematic errors not included here are: (1) 1$\sigma$ wavelength
errors of $\sim$0.76 \AA\ (due mainly to FGW errors) and (2) $\sim$10-20\%
errors in the EWs because of near neighbour and sky contamination (see text).}

\enddata
\end{deluxetable}





\begin{deluxetable}{lcccccccc}
\tablecolumns{9}
\tablewidth{0pc}
\tablecaption{Absolute magnitudes of CV secondaries and disks}
\tablehead{
\colhead{CV} & \colhead{$M_V$} & 
\colhead{$M_V$} & \colhead{Mass (\mdot)\tablenotemark{a}} & 
\colhead{predicted\tablenotemark{b}} & \colhead{$M_V$} & 
\colhead{$M_V$\tablenotemark{c}} & \colhead{$M_V$\tablenotemark{d}} &
 \colhead{$M_V$\tablenotemark{e}} \\
\colhead{} & \colhead{(total)} & 
\colhead{(secondary)} & \colhead{(secondary)} & 
\colhead{period (h)} & \colhead{(disk)} & \colhead{(disk)} &
\colhead{(disk)} & \colhead{(HeII)} \\
} 
\startdata

CV 1  & 5.95 & 5.93 & 0.68 & 5.1 &\nodata& 8.5 &  7.6  & 8.5 \nl
CV 2  & 7.13 & 7.24 & 0.57 & 4.4 &  9.7 &  8.7 &  8.2  & 8.5 \nl
CV 3  & 7.84 & 8.18 & 0.49 & 3.8 &  9.3 &  8.9 &  8.8  & 8.7 \nl
CV 4  & 8.41 & 8.53 & 0.45 & 3.6 & 10.9 &  9.9 &  9.9  & 9.3 \nl

\tablenotetext{a}{The secondary masses are from Bergbusch and Vandenberg
(1992) using the estimated $M_V$ values, and differ from those derived by
CG98, using different stellar evolution models, by at most only 0.01
\mdot.}

\tablenotetext{b}{Using Warner (1995) and Stehle, Kolb \& Ritter (1997).}

\tablenotetext{c}{After making the estimated $I$ magnitude of the secondary
fainter by 0.1 mag}

\tablenotetext{d}{After making the estimated $I$ magnitude of the secondary
fainter by 0.2 mag}

\tablenotetext{e}{Predicted $M_V$ using the HeII \la4686 line ratios given
in Section \ref{sec.cv4} and the \he2\ \la4686/$M_V$ correlation discussed
at the end of Section \ref{sec.discv}.}

\tablecomments{We have assumed an apparent distance modulus of 12.28 (from CG98)}

\enddata
\end{deluxetable}

\newpage

\vspace*{-4cm}

\begin{figure}[htb]
\vbox to8cm{\rule{0pt}{8cm}}
\includegraphics{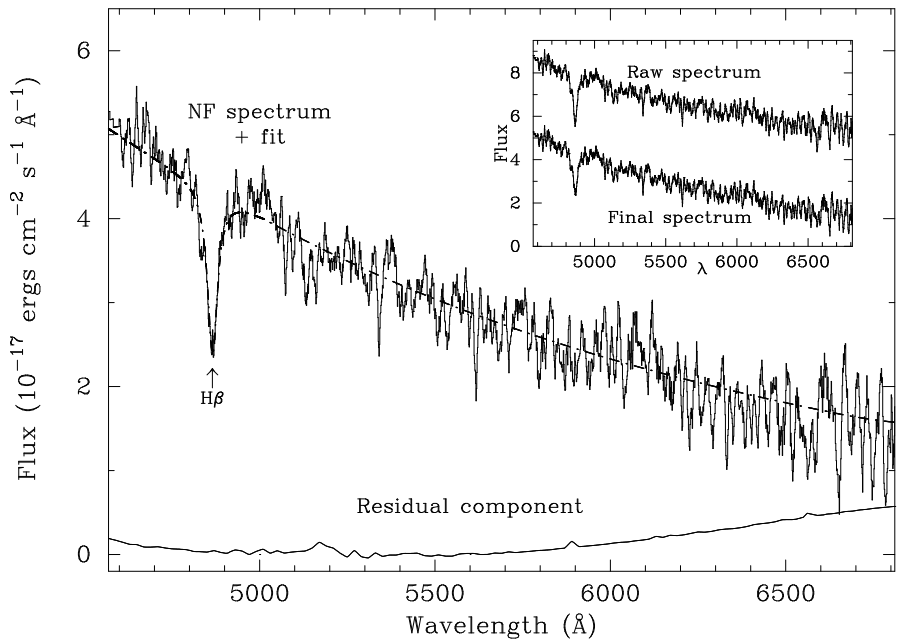}
\caption{The G570H spectrum of the NF, after normalizing its continuum
using the photometry of CG98. The model profile resulting from a fit to the
\hb\ line is also shown, plus the residual component described in the
text. The inset shows a comparison between the raw spectrum and the
normalized (``final'') spectrum. Note the small difference in the \hb\ line
profile between the raw and final spectrum because of the removal of the
spectral components from the neighboring star and diffuse light.} \label{fig.allnf}
\end{figure}

\clearpage
\pagebreak

\begin{figure}[htb]
\vbox to6cm{\rule{0pt}{6cm}}
\includegraphics{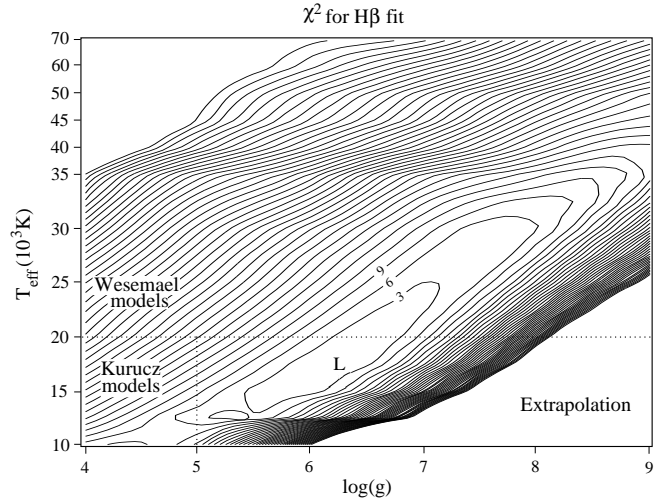}
\caption{A contour plot of $\chi^2$ for the fit of the Wesemael and Kurucz
line profiles to the \hb\ line of the NF. The ``L'' shows the optimal
solution and the dotted lines separate regions where the Wesemael and
Kurucz models have been used, and where they have been extrapolated. The
first ($\chi^2$=3), second ($\chi^2$=6) and fourth ($\chi^2$=12) contour
levels correspond roughly to 1$\sigma$, 2$\sigma$ and 3$\sigma$
respectively.} \label{fig.contnf} 
\end{figure}

\vspace*{4cm}

\begin{figure}[htb]
\vbox to3cm{\rule{0pt}{3cm}}
\includegraphics{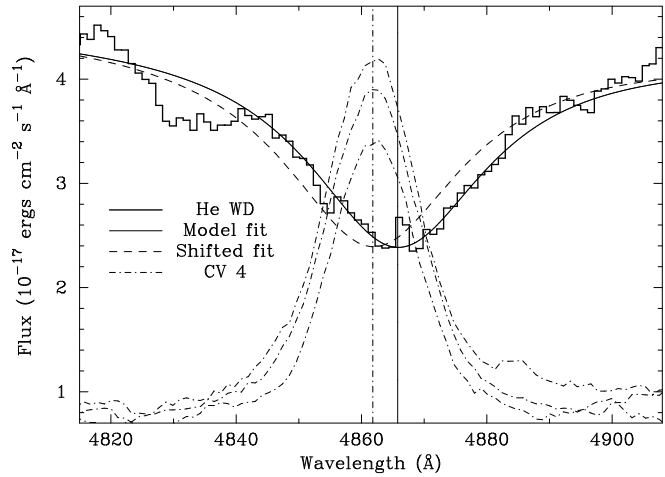}
\caption{A closeup of the \hb\ line profile of the NF. The noisy histogram
plots the data, the smooth solid line is a Lorentzian model fit to the
data, and the dot-dashed emission lines are from the three separate
observations of CV 4 (our wavelength reference), normalized to fit in our
figure. The dashed absorption line is the above Lorentzian fit but shifted
to the wavelength of the CV 4 line, and the vertical lines show the central
wavelengths of the NF and CV 4 \hb\ lines.} \label{fig.closenf}
\end{figure}

\begin{figure}[htb]
\vbox to12cm{\rule{0pt}{12cm}}
\includegraphics{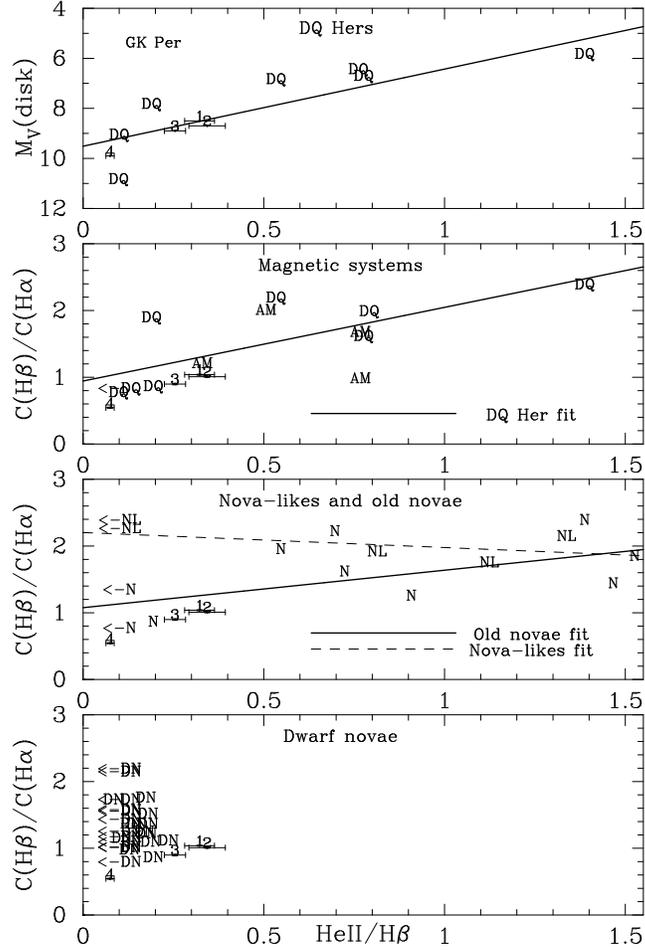}
\caption{Plots of the continuum ratio between \hb\ and H$\alpha$ versus the
\he2/\hb\ line ratio for CVs 1--4 along with several different classes of
field CV (from Williams 1983). $\chi^2$ fitting straight lines are shown
for individual CV classes where the linear correlation between the
continuum and line ratios has absolute value $>$ 0.5. Systems with
unmeasurable \he2\ are plotted at \he2/\hb\ = 0.1, and the $M_V$ (disk)
values are taken from column 7 of Table 2.} \label{fig.he2disk}
\end{figure}


\end{document}